\documentclass[sigconf,screen]{acmart}

\usepackage{microtype}

\usepackage{graphicx}

\usepackage{booktabs}
\usepackage{pgfplotstable} 
\usepackage{colortbl}
\usepackage{multirow}

\usepackage{paralist}

\usepackage[linesnumbered, ruled, vlined]{algorithm2e}
\SetAlFnt{\small}
\SetAlCapFnt{\small}
\SetAlCapNameFnt{\small}
\SetKwRepeat{Do}{do}{while}

\usepackage{amsthm}
\newtheoremstyle{custom}
  {0pt}   
  {-3pt}   
  {\normalfont}  
  {0pt}       
  {\bfseries} 
  {.}         
  {5pt plus 1pt minus 1pt} 
  {}
\theoremstyle{custom}

\usepackage{framed}

\usepackage{xcolor}
\usepackage{listings}
\definecolor{codegreen}{rgb}{0,0.6,0}
\definecolor{codegray}{rgb}{0.5,0.5,0.5}
\definecolor{codepurple}{rgb}{0.58,0,0.82}
\definecolor{backcolour}{rgb}{0.95,0.95,0.92}
\lstdefinestyle{mystyle}{
    backgroundcolor=\color{backcolour},   
    commentstyle=\color{codegreen},
    keywordstyle=\color{magenta},
    numberstyle=\tiny\color{codegray},
    stringstyle=\color{codepurple},
    basicstyle=\ttfamily\footnotesize,
    breakatwhitespace=false,         
    breaklines=true,                 
    captionpos=b,                    
    keepspaces=true,                 
    numbers=none,                    
    showspaces=false,                
    showstringspaces=false,
    showtabs=false,                  
    tabsize=2,
    frame=single,
    framesep=2pt,
    boxpos=c
}
\usepackage{tcolorbox}  

\usepackage{setspace}

\usepackage{multicol}  
\usepackage{subfigure}

\AtBeginDocument{%
  }

\setcopyright{cc}
\setcctype{by-nc-nd}
\acmDOI{10.1145/3832783.3834368}
\acmYear{2026}
\copyrightyear{2026}
\acmISBN{979-8-4007-2882-2/2026/10}
\acmConference[ASE '26]{Proceedings of the 41st IEEE/ACM International Conference on Automated Software Engineering}{October 12--16, 2026}{Munich, Germany}
\acmBooktitle{Proceedings of the 41st IEEE/ACM International Conference on Automated Software Engineering (ASE '26), October 12--16, 2026, Munich, Germany}
\acmSubmissionID{ase26main-p872-p}
\received{2026-03-26}
\received[accepted]{2026-06-18}

\begin{document}

\title{Evolution-Aware Regression Test Prioritization of ML-Enabled Systems using Gradient-Based Behavior Vectors}

\author{Eunho Cho}
\correspondingauthor
\orcid{0000-0002-4293-945X}
\affiliation{%
  \institution{KAIST}
  \department{School of Computing}
  \city{Daejeon}
  \country{Republic of Korea}
}
\email{ehcho@kaist.ac.kr}

\author{Donghwan Shin}
\orcid{0000-0002-0840-6449}
\affiliation{%
  \institution{University of Sheffield}
  \department{School of Computer Science}
  \city{Sheffield}
  \country{United Kingdom}
}
\email{d.shin@sheffield.ac.uk}

\author{In-Young Ko}
\orcid{0000-0002-3843-263X}
\affiliation{%
  \institution{KAIST}
  \city{Daejeon}
  \country{Republic of Korea}
}
\email{iko@kaist.ac.kr}

\renewcommand{\shortauthors}{Cho et al.}

\begin{abstract}
The machine learning~(ML) component of an ML-enabled system evolves through retraining, fine-tuning, and optimization, so previously valid test results may no longer hold. A single evolution step can worsen performance on some test cases while improving others, making regression test prioritization inherently directional. We present Gradient-based Behavior Vector-Parameter Delta~(GBV-PD), the first approach to operationalize the behavior vector space for evolution-aware regression test prioritization. GBV-PD represents each test case as a gradient-based vector~(GBV), a low-dimensional projection of its loss gradient under the original model. It then projects the observed parameter update of the evolved model onto the same PCA basis and uses the resulting alignment to estimate whether each test case's loss is likely to increase or decrease, without running the evolved model on test cases during prioritization. In an empirical study across classification and regression tasks, GBV-PD consistently outperformed non-directional baselines and remained competitive with a full-gradient reference, while offering better time and storage profiles for repeated updates via reusable GBV caching. These results show that behavior-space ideas can be operationalized into a practical and efficient mechanism for repeated-update regression-test prioritization of evolving ML components within ML-enabled systems.
\end{abstract}



\maketitle

\section{Introduction}

Many software systems are becoming ML-enabled, with their core functionality depending on machine learning~(ML) components~\cite{li2022testing}. This shift changes not only how such systems are built, but also how they should be tested~\cite{zhang2020machine}. In conventional software, behavior changes are usually tied to explicit code modifications, allowing regression testing to focus on test cases relevant to those modifications. In ML-enabled systems, however, behavior can change even when the architecture remains fixed, because model parameters may evolve through retraining, fine-tuning, optimization, or the incorporation of additional data~\cite{amershi2019software}. Such evolution can alter the model’s decision boundaries in ways that are not directly visible in the code, making it difficult to determine whether previously reliable test results still hold.

This challenge becomes especially important during maintenance. Under limited testing budgets, practitioners therefore face a familiar but challenging question: \textit{``Which tests should be rerun first after updating ML components?''} This is a well-known regression testing problem, but one complicated by the fact that the relevant behavioral changes are driven by parameter changes rather than explicit program edits~\cite{you2025navigating,li2022testing}. After ML model evolution, the evolved model may behave worse on some test cases and better on others, but these changes are not known in advance~\cite{amershi2019software,li2023lightweight}. An effective test prioritization approach must therefore account for the ``direction'' of the evolution--how the parameter updates--to distinguish regressions from improvements.

Existing approaches only partially address this issue. Black-box testing~\cite{zohdinasab2023efficient}, which relies only on inputs and outputs, provides limited insight into how the model has changed internally. Methods based on internal signals, such as neuron coverage~\cite{pei2017deepxplore} and surprise adequacy~\cite{kim2019guiding}, capture aspects of model responses. Still, they are not designed to prioritize tests based on the model evolution. Some also require executing the evolved model on a large test set, which becomes costly when updates are frequent in modern MLOps pipelines~\cite{kreuzberger2023machine,renggli2019continuous}, including parameter-efficient fine-tuning workflows such as LoRA~\cite{hu2022lora}.

To directly account for evolution, we formulate regression test prioritization as a directional ranking problem in a shared vector space, prioritizing test cases based on the alignment between each test-case vector and the observed parameter update $\Delta \theta$ as a vector.
We represent each test case as a Gradient-based Behavior Vector~(GBV), a low-dimensional vector derived from the gradient of the loss with respect to the original model parameters. These vectors form a Behavior Vector Space~(BVS), in which test cases are positioned not by surface-level input features but by their behavioral sensitivity to model changes~\cite{cho2026towards}.
Our key insight is that when a model's parameters change, the resulting shift in loss for a specific test case can be approximated by the inner product of the test case's GBV and the parameter update vector. This allows us to infer model behavior from the ``direction'' of the update--that is, whether a given parameter shift is likely to increase or decrease the loss for a specific test case--without execution of the evolved model.

We operationalize this idea into a novel method for regression test prioritization: GBV-Parameter-Delta~(GBV-PD). Our method precomputes the gradients for individual test cases once on the original model and compresses them into GBVs using Principal Component Analysis~(PCA). For any subsequent model update within the same architecture, it simply computes the parameter delta~(which is much cheaper than running the updated model on all test cases) and projects it onto the same PCA basis. The alignment between a test case's GBV and this projected update vector provides a lightweight estimate of performance shifts. By combining this directional estimate with the original loss, GBV-PD effectively prioritizes test cases so regressions can be detected earlier than might otherwise be possible. To the best of our knowledge, this is the first method to leverage BVS for efficient regression-test prioritization without executing the evolved model on the test suite in iterative ML development. 

The main contributions of this paper are as follows.
\begin{itemize}[-]
    \item A novel formulation of post-update regression test prioritization for evolving ML components as a directional ranking problem, which explicitly considers the ``direction'' of model updates to distinguish regressions from improvements.
    \item A novel prioritization method, GBV-PD, that operationalizes BVS for iterative ML development for the first time by projecting model parameter updates onto reusable GBVs, thereby enabling test prioritization without running the evolved model on test cases.
    \item An empirical evaluation across image classification and behavior-cloning tasks, with a replication package~\cite{replication}, demonstrating that GBV-PD consistently outperforms non-directional baselines, remains competitive with a full-gradient directional reference, and offers a more favorable time and storage profile under repeated updates.
\end{itemize}

\section{Problem Definition}
\label{sec:problem}

\textit{Evolution-Aware Regression Test Prioritization}.
We consider ML-enabled systems in which the core ML component evolves through parameter updates that preserve the architecture. Let $f_{\theta}$ denote the original model and $f_{\theta+\Delta\theta}$ the updated model. Let $T=\{t_i\}_{i=1}^{n}=\{(x_i, y_i)\}_{i=1}^{n}$ be a test set, where each test case $t_i$ consists of an input $x_i$ and its expected class label or target value $y_i$. The output of the prioritization procedure is a full ranking $\pi=(\pi_1,\ldots,\pi_n)$ over the tests, where $\pi_j$ is the $j$-th test case to execute. Under a limited testing budget that permits the execution of only $k \le n$ test cases, practitioners run the sequence of test cases $S_k=(t_{\pi_1},\ldots,t_{\pi_k})$ based on the top-$k$ prefix of ranking. The goal is therefore to construct a ranking whose early prefix exposes as many relevant post-evolution behavior changes as possible.

\textit{Regression and Improvement}.
After model evolution, the updated model may fail on some test cases that the original model handled correctly, and may handle correctly some test cases on which the original model previously failed. We refer to these as \emph{regression}~($\mathcal{R}$) and \emph{improvement}~($\mathcal{I}$) cases, respectively. For classification, let
\[
c_{\theta}(x,y)=\mathbf{1}[\arg\max f_{\theta}(x)=y],
\]
where $\mathbf{1}[\cdot]$ is the indicator function, which returns 1 when its argument is true and 0 otherwise. We assume that the output $f_\theta(x)$ of a classification model is not the class label itself, but a vector of class-wise scores or probabilities.
For each test $t_i=(x_i,y_i)$:
\[
t_i \in \mathcal{R}
\iff
c_{\theta}(x_i,y_i)=1
\ \wedge\
c_{\theta+\Delta\theta}(x_i,y_i)=0,
\]
\[
t_i \in \mathcal{I}
\iff
c_{\theta}(x_i,y_i)=0
\ \wedge\
c_{\theta+\Delta\theta}(x_i,y_i)=1.
\]

For regression tasks, let $e_{\theta}(x,y)$ be a task-specific error function and $\tau_{\mathrm{task}}>0$ a threshold for meaningful change:
\[
t_i \in \mathcal{R}
\iff
e_{\theta+\Delta\theta}(x_i,y_i)-e_{\theta}(x_i,y_i)>\tau_{\mathrm{task}},
\]
\[
t_i \in \mathcal{I}
\iff
e_{\theta}(x_i,y_i)-e_{\theta+\Delta\theta}(x_i,y_i)>\tau_{\mathrm{task}}.
\]
These definitions capture meaningful behavior changes of the model on a test case rather than small numerical fluctuations. The concrete error functions and thresholds are specified per task in Section~\ref{subsec:setup}. The resulting labels are available only after evaluation and cannot be used during prioritization.

\textit{Objective}.
The objective of this work is regression-oriented prioritization: rank tests so that members of $\mathcal{R}$ appear as early as possible. Although practitioners typically focus on potential regressions, we also consider improvement-oriented prioritization as a complementary view, since it is not known in advance which test cases the updated model will perform better or worse on. For a given budget $k$, the ideal rankings are
\[
\pi^{\mathrm{reg}}
=
\arg\max_{\pi}
\sum_{j=1}^{k}
\mathbf{1}[t_{\pi_j}\in\mathcal{R}],
\qquad
\pi^{\mathrm{imp}}
=
\arg\max_{\pi}
\sum_{j=1}^{k}
\mathbf{1}[t_{\pi_j}\in\mathcal{I}].
\]

Because $\mathcal{R}$ and $\mathcal{I}$ are unknown at prioritization time, we seek to approximate these rankings using only the original model and the parameter update, without evaluating the updated model on any test cases.

\textit{Assumptions}. We assume that (1) the original and evolved models share the same architecture with aligned parameters, as in retraining or fine-tuning; (2) each test case has a ground-truth label or target value, allowing both task-loss computation and post-hoc evaluation; (3) a differentiable task loss $\mathcal{L}(f_{\theta}(x), y)$, such as cross-entropy for classification or a regression loss for continuous outputs, and its gradient are available for the original model; and (4) the parameter update $\Delta \theta$ is small enough that a first-order approximation preserves the relative ordering of update effects across tests.

\textit{Design Requirements}.
The repeated-update scenario above imposes practical requirements on an effective prioritization method.

\begin{itemize}[-]
    \item \textit{Update Sensitivity}: it should be sensitive to $\Delta\theta$ and reflect how the current parameter update affects each test case, not only how difficult that case was for the original model.
    \item \textit{Compactness}: because per-test learning signals, such as gradients, are high-dimensional, the method should compress them into reusable representations.
    \item \textit{Efficiency}: expensive offline computations performed once on the original model should be reusable across future updates, so that each new update requires only lightweight online scoring.
    \item \textit{Automation}: scoring and ranking should rely solely on artifacts available at prioritization time--i.e., the original model, cached artifacts, and the observed update--without manual intervention or ad-hoc rules specific to updates.
\end{itemize}

These requirements motivate our gradient-based behavior representation and parameter-delta projection method, described in the next section.

\begin{figure*}[ht]
  \centering
  \vspace{-0mm}
  \includegraphics[width=\linewidth]{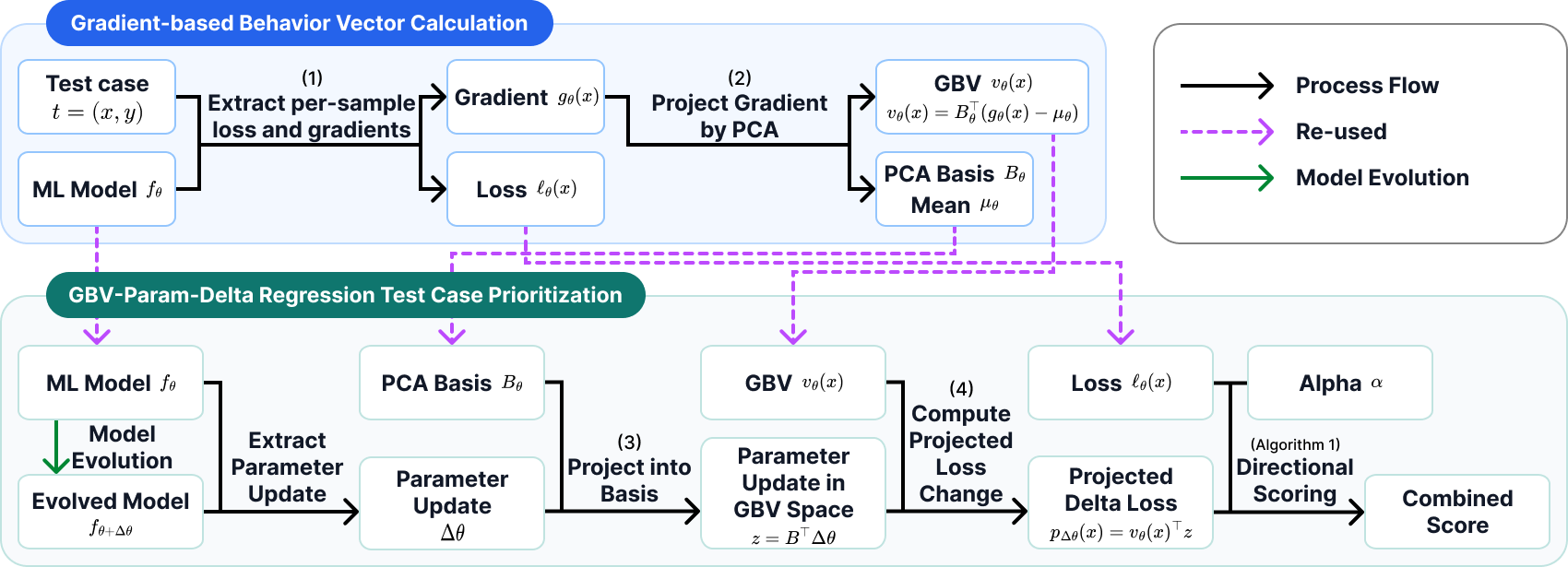}
  \caption{Overall workflow of GBV-PD. Offline, we compute per-test losses and gradients on the original model, learn a PCA basis, and cache GBVs. For each update, we express the parameter update in the same PCA coordinate system as the cached GBVs, use their inner products to estimate per-test projected delta loss, and rank tests for prioritization based on regression- or improvement-oriented criteria.}
  \label{fig:approach}
  \vspace{-0mm}
\end{figure*}

\section{Gradient-based Behavior Vector-Parameter Delta}
\label{sec:approach}

Figure~\ref{fig:approach} illustrates our gradient-based operationalization of BVS for evolution-aware test prioritization. The method addresses the design requirements by representing each test case compactly, reusing cached offline computation, and remaining sensitive to the current parameter update. Each test case is represented as a vector in a compact behavior subspace derived from its per-test gradient under the original model.

The key idea is to express the current model update in that same subspace. Because per-test gradients and parameter updates both lie in the target-layer parameter space, they can be compared in a shared coordinate system. Once both are written in the same PCA coordinates, their inner product provides a compact approximation of how the update is likely to change each test case's loss.

Prioritization, therefore, becomes directional: alignment with the update suggests a loss increase, whereas opposite alignment suggests a loss decrease. In practice, we operate on a target layer set $K$ rather than all parameters; accordingly, $g_\theta(x)$ and $\Delta\theta$ denote flattened quantities over $\theta[K]$.

\subsection{Gradient-Based Operationalization of Behavior Vector Space}

For a test case $t=(x,y)$ under the original model $f_\theta$, we define the loss $\ell_\theta$ and gradient $g_\theta$ for a test input $x$ as
\[
\ell_\theta(x)=\mathcal{L}(f_\theta(x), y), \quad
g_\theta(x)=\nabla_{\theta[K]} \mathcal{L}(f_\theta(x), y).\tag{1}
\]
Here, $\ell_\theta(x)$ is the task-specific differentiable loss used to derive gradients. Because $g_\theta(x)$ is high-dimensional, directly storing one gradient per test case is expensive. We therefore learn a PCA basis from per-test gradients computed on a sampled subset of training data under the original model and project each gradient into a low-dimensional subspace.

Let $\mu_\theta$ be the mean gradient and $B_\theta \in \mathbb{R}^{d_K \times m}$ the PCA basis with $m \ll d_K$. We define the GBV as
\[
v_\theta(x)=B_\theta^\top(g_\theta(x)-\mu_\theta).\tag{2}
\]

The resulting GBV is a compact representation of how the loss of test case $x$ locally responds to parameter perturbations in the target layer set. Caching GBVs once allows the same test representations to be reused across repeated updates. The choice of target layer, PCA dimension $m$, and PCA sampling size trades off fidelity against cost; we study this trade-off in Section~\ref{subsec:rq4}.

\subsection{Parameter Delta Projection}

\subsubsection{Project Parameter Update into Basis}

A GBV stores each test case in PCA coordinates, whereas model evolution is observed as the parameter update $\Delta\theta$ in the original parameter space. To relate a cached GBV to the current update, we must express the update in the same coordinates; otherwise, the GBV and the update are not directly comparable. To use GBVs rather than full gradients, we project the evolution onto the same PCA basis used for GBV:
\[
z=B_\theta^\top\Delta\theta.\tag{3}
\]

\subsubsection{Compute Projected Delta Loss}

Then, our goal is to estimate the post-evolution loss change $\Delta \ell_\theta(x)$ using GBV and parameter updates in GBV space $z$. Let us express $\Delta \ell_\theta(x)$ as a function of GBV and $z$. First, let the post-evolution loss change for test input $x$ be
\[
\Delta\ell_\theta(x)=
\mathcal{L}(f_{\theta+\Delta\theta}(x), y)-\mathcal{L}(f_\theta(x), y).
\]

Treating $\mathcal{L}(f_\theta(x), y)$ as a function of $\theta$, a first-order Taylor expansion at $\theta$~\cite{boyd2004convex} gives
\[
\mathcal{L}(f_{\theta+\Delta\theta}(x), y)
\approx
\mathcal{L}(f_\theta(x), y)+g_\theta(x)^\top\Delta\theta,
\]
and therefore
\[
\Delta\ell_\theta(x)\approx g_\theta(x)^\top\Delta\theta.
\]

The post-evolution loss change $\Delta \ell_\theta(x)$ is thus estimated by the inner product between the gradient and the evolution. We decompose the centered gradient into its PCA-subspace component and residual:
\[
g_\theta(x)-\mu_\theta = B_\theta v_\theta(x)+r_\theta(x).
\]
Substituting this decomposition by GBV, and $z$, we obtain
\[
g_\theta(x)^\top\Delta\theta
=
\mu_\theta^\top\Delta\theta
+
v_\theta(x)^\top z
+
r_\theta(x)^\top (I-B_\theta B_\theta^\top)\Delta\theta.
\]

Since $\mu_\theta^\top\Delta\theta$ is constant across tests, it does not affect ranking. Ignoring the residual term for relative prioritization, we define the projected delta loss for each test $t_i=(x_i,y_i)$ as
\[
p_i \equiv p_{\Delta\theta}(x_i)=v_\theta(x_i)^\top z.\tag{4}
\]

This quantity is large when the current update points in a similar direction to the test case's gradient response in GBV space, and small or negative when it points in the opposite direction. Accordingly, $p_i$ is interpreted as a relative directional ranking score. Larger values indicate that the update is predicted to increase the test loss more—or decrease it less—than for other tests, whereas smaller values indicate the opposite.

\subsubsection{Projection Error of Delta Loss}

The accuracy of this proxy depends on two approximations. First, the Taylor approximation is more accurate when the loss surface is smooth near $\theta$ and the update magnitude is small. If the Hessian spectral norm, which represents the curvature of the loss along the path from $\theta$ to $\theta+\Delta\theta$ is bounded by a constant $\beta$, then the first-order approximation error $\epsilon_{\mathrm{taylor}}(x)$ satisfies~\cite{nocedal2006numerical}
\[
|\epsilon_{\mathrm{taylor}}(x)|
\le
\frac{\beta}{2}||\Delta\theta||_2^2.
\]

Second, the projection step is more accurate when the PCA subspace captures most of the relevant gradient and update mass. The projection error $\epsilon_{\mathrm{projection}}(x)$ satisfies
\[
|\epsilon_{\mathrm{projection}}(x)|
\le
||(I-B_\theta B_\theta^\top)(g_\theta(x)-\mu_\theta)||_2
\cdot
||(I-B_\theta B_\theta^\top)\Delta\theta||_2.
\]
These bounds are used only to justify relative ranking, not to predict exact post-update loss. When the update is not too large and the PCA subspace is faithful, $p_i$ preserves the relative directional differences needed for prioritization.

\subsection{GBV–Parameter-Delta Prioritization}

\begin{algorithm}[t]
\small
\caption{GBV-Parameter-Delta~(GBV-PD) for evolution-aware regression test prioritization}
\label{alg:param_delta}
\DontPrintSemicolon
\setstretch{0.85}
\SetKwInOut{Input}{Input}
\SetKwInOut{Output}{Output}

\Input{
    Original model parameters $\theta$; \\
    Evolved model parameters $\theta'$; \\
    GBV basis $B_\theta$; Test GBVs $V_\theta=\{v_\theta(x_i)\}_{i=1}^{n}$; \\
    Original losses $\ell_\theta=\{\ell_\theta(x_i)\}_{i=1}^{n}$; \\
    Mode $mode \in \{\texttt{reg}, \texttt{imp}\}$; Weight $\alpha$
}
\Output{Prioritized ranking $\pi$}

\BlankLine

$\mathcal{K} \leftarrow \textsc{GetTargetParameterKeys}(\theta,\theta')$\;\label{line:gbv_basis_start}
$\Delta\theta \leftarrow \textsc{Flatten}(\theta'[\mathcal{K}] - \theta[\mathcal{K}])$\;
$z \leftarrow B_\theta^\top \Delta\theta$\;\label{line:gbv_basis_end}

\For{$i \leftarrow 1$ \KwTo $n$}{\label{line:loss_projection_start}
    $p_i \leftarrow v_\theta(x_i)^\top z$\;
}\label{line:loss_projection_end}

$r^{(\ell)} \leftarrow \textsc{RankNormalizeDescending}(\ell_\theta)$\;\label{line:normalization_start}
$r^{(+)} \leftarrow \textsc{RankNormalizeDescending}(p)$\;
$r^{(-)} \leftarrow \textsc{RankNormalizeDescending}(-p)$\;\label{line:normalization_end}

\If{$mode = \texttt{reg}$}{\label{line:scoring_start}
    $s \leftarrow r^{(\ell)} + \alpha \, r^{(+)}$\;
}
\Else{
    $s \leftarrow r^{(\ell)} + \alpha \, r^{(-)}$\;
}\label{line:scoring_end}

$\pi \leftarrow \textsc{ArgsortDescending}(s)$\;\label{line:sorting}
\Return{$\pi$}\;\label{line:selection}
\end{algorithm}

We combine the original loss with the projected delta loss to rank test cases. Algorithm~\ref{alg:param_delta} presents the GBV-Parameter-Delta~(GBV-PD) procedure. Given cached GBVs $V_\theta$, original losses $\ell_\theta$, and the PCA basis, we first extract the target-layer parameter delta and project it into GBV space, producing $z=B_\theta^\top\Delta\theta$~(lines~\ref{line:gbv_basis_start}--\ref{line:gbv_basis_end}). For each test case $t_i=(x_i,y_i)$, we then compute the projected delta loss $p_i=v_\theta(x_i)^\top z$~(lines~\ref{line:loss_projection_start}--\ref{line:loss_projection_end}).

The original loss and the projected delta loss play different roles. $\ell_\theta(x_i)$ reflects how the original model already handles test case $t_i$, whereas $p_i$ reflects how the update is likely to move that behavior.

We rank-normalize these signals so that they can be combined despite their different scales and distributions. For regression- or improvement-oriented prioritization, larger projected increases or decreases in loss should receive higher priority. Accordingly, $r_i^{(+)}$ and $r_i^{(-)}$ are the descending rank of $p_i$ and $-p_i$, respectively~(lines~\ref{line:normalization_start}--\ref{line:normalization_end}). The parameter $\alpha$ controls how strongly the current update direction influences the final ranking relative to the original loss~(lines~\ref{line:scoring_start}--\ref{line:scoring_end}).

GBV-PD outputs a full ranking $\pi$~(lines~\ref{line:sorting}--\ref{line:selection}); under a budget $k$, practitioners execute its top-$k$ prefix. During prioritization, the procedure uses only cached GBVs, cached losses, and the current parameter delta. This approach does not require evaluating the updated model on any test cases.

\subsection{Computational Cost}

GBV-PD separates one-time offline costs from per-update online costs. Offline, we compute per-test losses and gradients, learn the PCA basis, and cache $\ell_\theta$, $V_\theta$, and $B_\theta$. Online, given an update, we form $\Delta\theta$, compute $z=B_\theta^\top\Delta\theta$, evaluate $p_i=v_\theta(x_i)^\top z$ for all tests, and sort the resulting scores.

Let $n$ denote the number of tests, $d_K$ the gradient dimension of the target layer, and $m$ the GBV dimension. The dominant online costs are $O(d_K)$ to extract $\Delta\theta$, $O(md_K)$ to compute $z$, $O(nm)$ to score all tests, and $O(n\log n)$ to rank them. Storage is $O(nm + md_K)$ for GBV caching, compared with $O(nd_K)$ for full-gradient caching. Since $m \ll d_K$ in typical settings, the cache is substantially smaller.

\section{Evaluation}
\label{sec:evaluation}

We evaluate GBV-PD along four dimensions:
\begin{enumerate}[\bf RQ1]
	\item \textbf{Validity}. How well does the GBV-based projected signal approximate the ranking of update-induced loss changes?
	\item \textbf{Effectiveness}. How effectively does GBV-PD place regression and improvement cases near the head of the ranking, compared with representative baselines?
	\item \textbf{Efficiency}. How practical is GBV-PD's cost structure in repeated-update settings?
	\item \textbf{Robustness}. How robust is GBV-PD to changes in behavior vector construction and scoring choices?
\end{enumerate}

GBV-PD's ranking is built from a projected loss-change signal. We check whether that signal is a reasonable proxy for the actual update-induced loss change in RQ1. RQ2 is the primary practical question: whether the ranking helps post-update testing by prioritizing regression and improvement cases early under budget. RQ3 and RQ4 then assess whether the method is deployable and stable in the repeated-update setting from Section~\ref{sec:problem}.

\subsection{Experiment Setup}
\label{subsec:setup}

\begin{table}[t]
\centering
\small
\caption{Experiment settings used in the evaluation. Rows report the evolution types, GBV configurations, and numbers of models/configurations for each task--dataset--architecture setting.}
\label{table:setup}
\begin{tabular}{|>{\centering\arraybackslash}m{1.6cm}|
                >{\centering\arraybackslash}m{1.3cm}|
                >{\centering\arraybackslash}m{1.4cm}|
                >{\centering\arraybackslash}m{1.2cm}|
                >{\centering\arraybackslash}m{1.2cm}|}
\hline
\textbf{Task} & \multicolumn{3}{c|}{Image Classification} & Behavior Cloning \\ \hline
\textbf{Dataset} & \multicolumn{3}{c|}{CIFAR100} & Udacity Jungle \\ \hline
\textbf{Architecture} & ViT-Tiny & ResNet18 & \multicolumn{2}{c|}{LeNet5} \\ \hline
\textbf{Evolution} & \multicolumn{3}{c|}{Global, Class-specific} & Global \\ \hline
\textbf{GBV Dim} & 64, 128, 256 & 128, 256, 512 & 32, 64, 128 & 16, 32, 64 \\ \hline
\textbf{Gradient Layer} & \multicolumn{2}{c|}{d} & \multicolumn{2}{c|}{abcd, d} \\ \hline
\textbf{\# Original Models} & 50 & 50 & 50 & 50 \\ \hline
\textbf{\# Evolution / Model} & 3 & 3 & 3 & 1 \\ \hline
\textbf{\# GBV Configs} & 12 & 12 & 24 & 24 \\ \hline
\end{tabular}
\end{table}

\subsubsection{Tasks, Datasets, and Model Architectures}
We evaluate two tasks: image classification~(IC) on the CIFAR-100~\cite{krizhevsky2009learning} and behavior cloning~(BC) on the Udacity Jungle dataset~\cite{su2023Udacity}, released as part of Udacity's behavior cloning project~\cite{Udacity2016CarND}. We chose two complementary supervised tasks with different prediction targets: classification for IC and continuous regression for BC. Together, these tasks let us assess whether the method remains useful beyond classification while staying in a label-available, same-architecture setting. Because the Udacity dataset has no predefined split, we use the first 80\% of instances for training and the last 20\% for testing by index. For behavior cloning, we treat changes in prediction with an absolute magnitude of at most $10^{-5}$ as unchanged to avoid counting numerically negligible differences as regressions or improvements. For IC, we evaluate LeNet5~\cite{lecun1998gradient}, ResNet18~\cite{he2016deep}, and ViT-Tiny~\cite{dosovitskiy2021image}; for BC, we use LeNet5.

\subsubsection{Target Models}

For each setting, we construct 50 original models by varying the random seed~(42--46), training epoch~(0, 25, 50, 100), and, for nonzero epochs, learning rate~($5\times10^{-3}$, $10^{-3}$, $5\times10^{-4}$). All models are trained with AdamW on the full training set without data augmentation. We consider two evolution types. A \textit{global update} trains the model for one additional epoch on the full training set with a learning rate of $10^{-5}$. A \textit{class-specific update}, used only for CIFAR100, trains for one additional epoch with a learning rate of $10^{-4}$ on data from a single superclass~(\texttt{vehicle\_1} or \texttt{large\_carnivores}), letting us study both global and localized behavior changes. For behavior cloning, we use only the global update. These updates are deliberately mild and structure-preserving: they reflect maintenance-style retraining or fine-tuning while still producing mixed regression and improvement cases. This is the kind of evolution targeted by the first-order approximation used by GBV-PD.

\subsubsection{Testing Configurations}

Table~\ref{table:setup} lists the GBV dimensions and gradient-layer settings. Here, \texttt{d} uses only the final fully connected layer, whereas \texttt{abcd} uses one layer each from the top, middle, bottom, and final fully connected layers. PCA-basis construction combines two initial-sample choices with two sample sizes, yielding four sampling settings. Unless otherwise noted, we evaluate budgets $k \in \{10,20,50,100,200,500\}$ and use $\alpha \in \{1,2,3\}$ for GBV-PD and Grad-PD; RQ4 additionally considers $\alpha \in \{0,0.5\}$.

\subsubsection{Baselines}

Few published methods target exactly the same goal as we do---post-update prioritization for repeated same-archite-cture updates---so we compare against three baseline families. \textbf{Random} and \textbf{Loss-only} are simple non-directional references. \textbf{Neuron Coverage}~(NC)~\cite{pei2017deepxplore}, \textbf{Surprise Adequacy}~(SA)~\cite{kim2019guiding}, and \textbf{TDPR}~\cite{shen2024prioritizing} are stronger model-aware ranking heuristics, but none use the current parameter update; they therefore test whether generic difficulty, internal-response, or training-dynamics signals are sufficient. \textbf{MetaSel}~(MS)~\cite{abbasishahkoo2025metasel} is the closest prior method for evolving models, but it was proposed for test selection rather than ranking. To compare it fairly with GBV-PD's prioritization output, we convert its per-test score into a ranking and evaluate its top-$k$ prefixes under the same budgets and paired statistical protocol. Finally, \textbf{Grad-PD} is a directional full-information reference that uses the same parameter-delta idea without GBV compression. This comparison isolates the value of explicit update-direction modeling, and then the additional value of compression. Because MS and TDPR are defined only for classification tasks, we evaluate them only on image classification tasks.

\subsubsection{Reproducibility and Additional Results}

We implement the approach in Python 3.13 with PyTorch and CUDA and run experiments on Ubuntu 22.04 with an Intel Xeon Gold 4215R, 128~GB RAM, and an NVIDIA RTX A5000. All results are reproducible with the accompanying replication kit~\cite{replication}, which also contains omitted figures and full results.

\subsection{RQ1. Validity}

RQ1 is a prerequisite validity check on the ranking signal used by GBV-PD. For each test case $t_i=(x_i,y_i)$, we compute the actual post-update loss change $\Delta \ell_\theta(x_i)$ and the projected signal $p_i$, then measure their Spearman rank correlation.

To study sensitivity to update magnitude, we also evaluate scaled updates $\lambda \Delta\theta$ for $\lambda \in \{1,2,5,10,20\}$. Here, $\lambda=1$ corresponds to the actual maintenance-style updates used in the rest of the evaluation. In contrast, larger $\lambda$ values are diagnostic stress tests that progressively move the update away from the local regime assumed by the first-order approximation.

\begin{figure}[ht]
  \centering
  \vspace{-0mm}
  \includegraphics[width=0.9\linewidth]{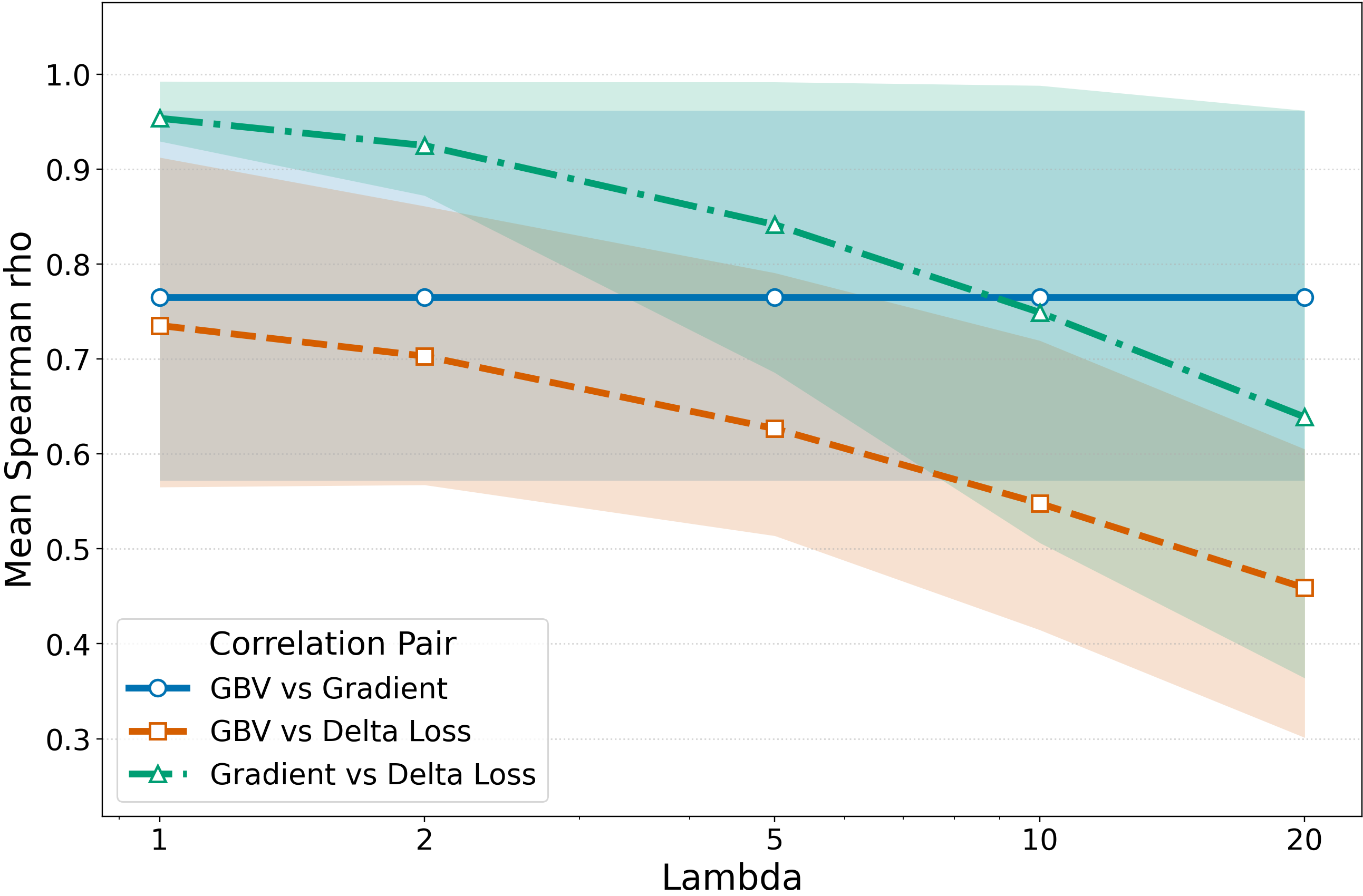}
  \caption{Spearman rank correlations across scaled update magnitudes $\lambda$ for a representative subject, IC~(LeNet5). The GBV-vs.-gradient correlation stays nearly flat, whereas both projected signals lose correlation with the true loss change as the update grows.}
  \label{fig:rq1}
  \vspace{-0mm}
\end{figure}

Figure~\ref{fig:rq1} shows a representative result; full subject-wise plots are included in the replication kit. At $\lambda=1$, the GBV projection remains strongly correlated with the true $\Delta \ell_\theta$ (about $0.74$), indicating that the compressed representation still captures which tests are most affected by the update. As $\lambda$ increases, both the full-gradient and GBV projections become less correlated with the true loss change.

The GBV-vs.-gradient curve stays nearly flat across $\lambda$, indicating that PCA compression itself is stable across update magnitudes. By contrast, both projected signals lose correlation with the true $\Delta \ell_\theta$ as $\lambda$ increases, consistent with $\epsilon_{\mathrm{taylor}}$: as the evolution grows, the first-order expansion becomes less faithful. This pattern is also consistent with $\epsilon_{\mathrm{projection}}$: the main degradation comes from the first-order expansion rather than from instability in the compression.

\begin{tcolorbox}
    In response to RQ1, the GBV-based projected signal is a useful proxy for ranking test cases by post-update loss change. It tracks much of the full-gradient directional signal in a low-dimensional space, and its validity is strongest in the low-magnitude evolution targeted by this paper. This result supports the claim that GBV-PD meets \textit{Update Sensitivity} requirement.
\end{tcolorbox}

\subsection{RQ2. Effectiveness}

RQ2 asks how effective GBV-PD is for prioritization. For mode $m \in \{\mathrm{reg}, \mathrm{imp}\}$, let $M_m$ be the corresponding target set, and let $S_k=(t_{\pi_1},\ldots,t_{\pi_k})$ denote the top-$k$ prefix of a method's ranking $\pi$. We report Precision@k, Recall@k, and nDCG@k using their standard binary-relevance definitions, together with Enrichment Ratio~(ER):
\[
\mathrm{ER}^{m}@k
=
\frac{|S_k \cap M_m|/k}{|Z_z|/n},
\]
where $n$ is the test-set size. ER measures how concentrated a ranking prefix is in target cases relative to the base rate. Among these metrics, nDCG@k is order-sensitive within the top-$k$ prefix and therefore captures prioritization quality beyond mere inclusion.

To summarize performance across budgets, we report the area under each metric curve on a log-scaled budget axis~(\textit{Metric-AUC}). We prefix metrics with R- and I- for regression- and improvement-oriented evaluation.

For each subject, the original trained model is the unit of paired analysis. Metric-AUC values from configurations associated with the same original model are first averaged, and only matched model-level observations are compared between GBV-PD and each baseline. This avoids treating configurations derived from the same trained model as independent samples--i.e., pseudo-replication--which would inflate the effective sample size and understate variance. We then apply two-sided Wilcoxon signed-rank tests with Holm correction.

\begin{table*}[ht]
\caption{Average Metric-AUC over top-$k$ prefixes for RQ2 across all subjects. Higher is better. In baseline cells, superscripts compare against GBV-PD using paired Wilcoxon signed-rank tests with Holm correction~(no superscript: GBV-PD is significantly better; $^{*}$, $^{**}$, and $^{***}$: the baseline is significantly better at corrected $p<0.05$, $p<0.01$, and $p<0.001$, respectively; $^{\sim}$ and $^{\approx}$: GBV-PD or the baseline is numerically better, respectively, without statistical significance). In GBV-PD cells, subscripts report the number of original models on which GBV-PD outperformed Grad-PD. MS and TDPR are not applicable to BC~(LeNet5).}
\label{table:auc}
\centering
\footnotesize

\resizebox{\textwidth}{!}{%

\begin{filecontents*}{table/02_auc.csv}
subject,metric,random,neuron_coverage,surprise_adequacy,loss_only,metasel,tdpr,gradi,gbvpdi,metric_r,random_r,neuron_coverage_r,surprise_adequacy_r,loss_only_r,metasel_r,tdpr_r,gradr_r,gbvpdr_r
,Precision,0.03,0.03,0.02,0,0.06,0.02,0.25,$0.28_{\scriptscriptstyle 31/50}$,Precision,0.02,0.02,0.02,0,0.06,0.02,$^{\scriptscriptstyle ***}0.29$,$0.29_{\scriptscriptstyle 16/50}$
IC,Recall,0.02,0.02,0.02,0,0.03,0.02,$^{\scriptscriptstyle ***}0.14$,$0.14_{\scriptscriptstyle 16/50}$,Recall,0.02,0.02,0.02,0,0.03,0.02,$^{\scriptscriptstyle ***}0.12$,$0.11_{\scriptscriptstyle 10/50}$
(ViT-Tiny),nDCG,0.03,0.03,0.03,0,0.05,0.02,$^{\scriptscriptstyle \sim}0.26$,$0.29_{\scriptscriptstyle 30/50}$,nDCG,0.03,0.03,0.02,0,0.06,0.02,$^{\scriptscriptstyle ***}0.37$,$0.35_{\scriptscriptstyle 15/50}$
,ER,1.08,1.21,0.98,0,2.19,0.9,$^{\scriptscriptstyle \sim}10.41$,$11.18_{\scriptscriptstyle 29/50}$,ER,1.06,1.06,0.93,0,2.52,1,$^{\scriptscriptstyle ***}13.72$,$12.53_{\scriptscriptstyle 14/50}$
,Precision,0.07,0.04,0.03,0,0.18,0.06,0.07,$0.10_{\scriptscriptstyle 49/50}$,Precision,0.08,0.07,0.07,0,$^{\scriptscriptstyle **}0.40$,0.07,0.32,$0.33_{\scriptscriptstyle 44/50}$
IC,Recall,0.02,0.02,0.02,0,0.05,0.02,0.05,$0.06_{\scriptscriptstyle 49/50}$,Recall,0.02,0.02,0.02,0,$^{\scriptscriptstyle *}0.07$,0.02,0.05,$0.06_{\scriptscriptstyle 48/50}$
(ResNet18),nDCG,0.06,0.04,0.03,0,0.17,0.06,0.07,$0.10_{\scriptscriptstyle 42/50}$,nDCG,0.07,0.06,0.06,0,$^{\scriptscriptstyle \approx}0.40$,0.06,0.31,$0.36_{\scriptscriptstyle 46/50}$
,ER,1.39,1.03,0.79,0,3.96,1.5,1.71,$2.68_{\scriptscriptstyle 46/50}$,ER,1.37,1.22,1.18,0,5.46,1.15,5.39,$6.21_{\scriptscriptstyle 48/50}$
,Precision,0.03,0.03,0.03,0,0.08,0.03,$^{\scriptscriptstyle \approx}0.17$,$0.17_{\scriptscriptstyle 20/50}$,Precision,0.03,0.03,0.03,0,0.2,0.03,$^{\scriptscriptstyle ***}0.56$,$0.47_{\scriptscriptstyle 0/49}$
IC,Recall,0.02,0.02,0.02,0,0.04,0.02,$^{\scriptscriptstyle ***}0.12$,$0.11_{\scriptscriptstyle 14/50}$,Recall,0.02,0.02,0.02,0,0.04,0.02,$^{\scriptscriptstyle ***}0.16$,$0.14_{\scriptscriptstyle 0/49}$
(LeNet5),nDCG,0.03,0.03,0.03,0,0.08,0.03,$^{\scriptscriptstyle **}0.20$,$0.18_{\scriptscriptstyle 17/50}$,nDCG,0.04,0.04,0.03,0,0.2,0.03,$^{\scriptscriptstyle ***}0.67$,$0.54_{\scriptscriptstyle 0/49}$
,ER,1.11,1.17,0.92,0,2.49,0.89,$^{\scriptscriptstyle ***}7.05$,$6.67_{\scriptscriptstyle 16/50}$,ER,1.11,1.07,0.84,0,3.42,0.92,$^{\scriptscriptstyle ***}14.88$,$12.42_{\scriptscriptstyle 0/49}$
,Precision,0.75,0.81,0.77,1,NaN,NaN,$^{\scriptscriptstyle \approx}1.49$,$1.54_{\scriptscriptstyle 25/50}$,Precision,0.67,0.75,0.67,0.69,NaN,NaN,$^{\scriptscriptstyle \approx}1.47$,$1.48_{\scriptscriptstyle 18/50}$
BC,Recall,0.33,0.34,0.33,0.37,NaN,NaN,$^{\scriptscriptstyle \approx}0.50$,$0.51_{\scriptscriptstyle 20/50}$,Recall,0.33,0.34,0.34,0.36,NaN,NaN,$^{\scriptscriptstyle \approx}0.55$,$0.55_{\scriptscriptstyle 17/50}$
(LeNet5),nDCG,0.79,0.84,0.81,1.09,NaN,NaN,$^{\scriptscriptstyle \approx}1.58$,$1.62_{\scriptscriptstyle 24/50}$,nDCG,0.7,0.74,0.72,0.7,NaN,NaN,$^{\scriptscriptstyle \approx}1.59$,$1.60_{\scriptscriptstyle 17/50}$
,ER,1.68,1.77,1.64,2.07,NaN,NaN,$^{\scriptscriptstyle \approx}2.79$,$2.92_{\scriptscriptstyle 25/50}$,ER,1.7,1.84,1.64,1.6,NaN,NaN,$^{\scriptscriptstyle \approx}3.22$,$3.27_{\scriptscriptstyle 18/50}$
\end{filecontents*}
\pgfplotstabletypeset[
    col sep=comma,
    every head row/.style={
        before row={\toprule},
        after row=\midrule
    },
    every nth row={4[-3]}{
        before row={\cmidrule(l){2-19}},
        after row={\cmidrule(l){2-19}},
    },
    every nth row={4[-1]}{
        before row={\cmidrule(l){2-19}},
        after row={\midrule},
    },
    every last row/.style={after row=\bottomrule},
    columns/subject/.style={
        column type=c,
        column name=Subject,
        string type
    },
    columns/metric/.style={column type=c, column name=I-Metric, string type},
    columns/random/.style={column type=r, column name=Rand, fixed, fixed zerofill, precision=2},
    columns/loss_only/.style={column type=r, column name=Loss, fixed, fixed zerofill, precision=2},
    columns/neuron_coverage/.style={column type=r, column name=NC, fixed, fixed zerofill, precision=2},
    columns/surprise_adequacy/.style={column type=r, column name=SA, fixed, fixed zerofill, precision=2},
    columns/metasel/.style={column type=r, column name=MS, string type},
    columns/tdpr/.style={column type=r, column name=TDPR, fixed, fixed zerofill, precision=2},
    columns/gradi/.style={column type=r, column name=Grad-PD, string type},
    columns/gbvpdi/.style={column type=r, column name=GBV-PD, string type},
    columns/metric_r/.style={column type=c, column name=R-Metric, string type},
    columns/random_r/.style={column type=r, column name=Rand, fixed, fixed zerofill, precision=2},
    columns/loss_only_r/.style={column type=r, column name=Loss, fixed, fixed zerofill, precision=2},
    columns/neuron_coverage_r/.style={column type=r, column name=NC, fixed, fixed zerofill, precision=2},
    columns/surprise_adequacy_r/.style={column type=r, column name=SA, fixed, fixed zerofill, precision=2},
    columns/metasel_r/.style={column type=r, column name=MS, string type},
    columns/tdpr_r/.style={column type=r, column name=TDPR, fixed, fixed zerofill, precision=2},
    columns/gradr_r/.style={column type=r, column name=Grad-PD, string type},
    columns/gbvpdr_r/.style={column type=r, column name=GBV-PD, string type}
]{table/02_auc.csv}%

}

\vspace{-0mm}
\end{table*}

\begin{figure}[ht]
  \centering
  \vspace{-0mm}
  \includegraphics[width=\linewidth]{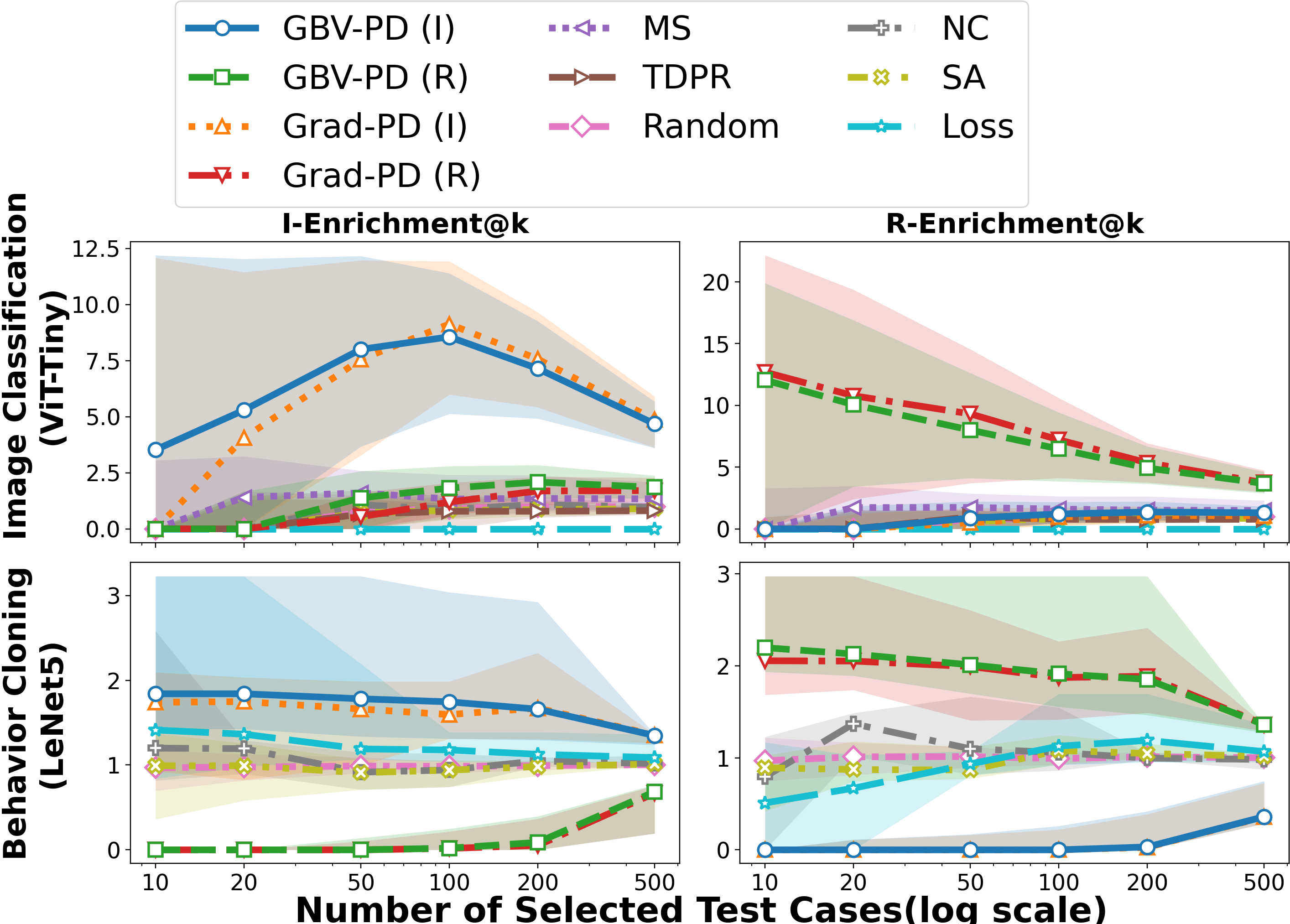}
  \caption{Representative I- and R-ER@k curves over log-scaled budgets for IC~(ViT-Tiny)~(top) and BC~(LeNet5)~(bottom). For Grad-PD and GBV-PD, (I) and (R) denote improvement- and regression-oriented rankings. Directional methods outperform non-directional baselines most clearly at small budgets.}
  \label{fig:rq2_1}
  \vspace{-0mm}
\end{figure}

Table~\ref{table:auc} summarizes the average Metric-AUC over all subjects, and Figure~\ref{fig:rq2_1} shows representative ER@k curves. MS and TDPR do not apply to the behavior-cloning subject; those cells are reported as NaN and excluded from the corresponding paired tests.

Overall, GBV-PD is one of the strongest practical prioritizers among the compared methods. Across subjects, it consistently outperforms the non-directional baselines, usually outperforms MS, and remains broadly competitive with Grad-PD. The advantage is clearest at small budgets, where prioritization matters most, as shown in Figure~\ref{fig:rq2_1}, and both GBV-PD modes move target cases much closer to the top of the ranking. Paired analysis at the original-model level confirms significant improvements over the applicable non-directional baselines after Holm correction.

This result matters because it isolates the value of evolution-direction awareness. Random, Loss-only, NC, SA, and TDPR provide signals about difficulty, activation patterns, surprise, or training dynamics, but none explicitly model how the \emph{current} update changes each test case. The consistent gap, therefore, suggests that the missing ingredient is not simply a better original-model heuristic, but explicit alignment between per-test behavior and the observed update direction.

\begin{figure}[t]
  \centering
  \vspace{-0mm}
  \includegraphics[width=0.95\linewidth]{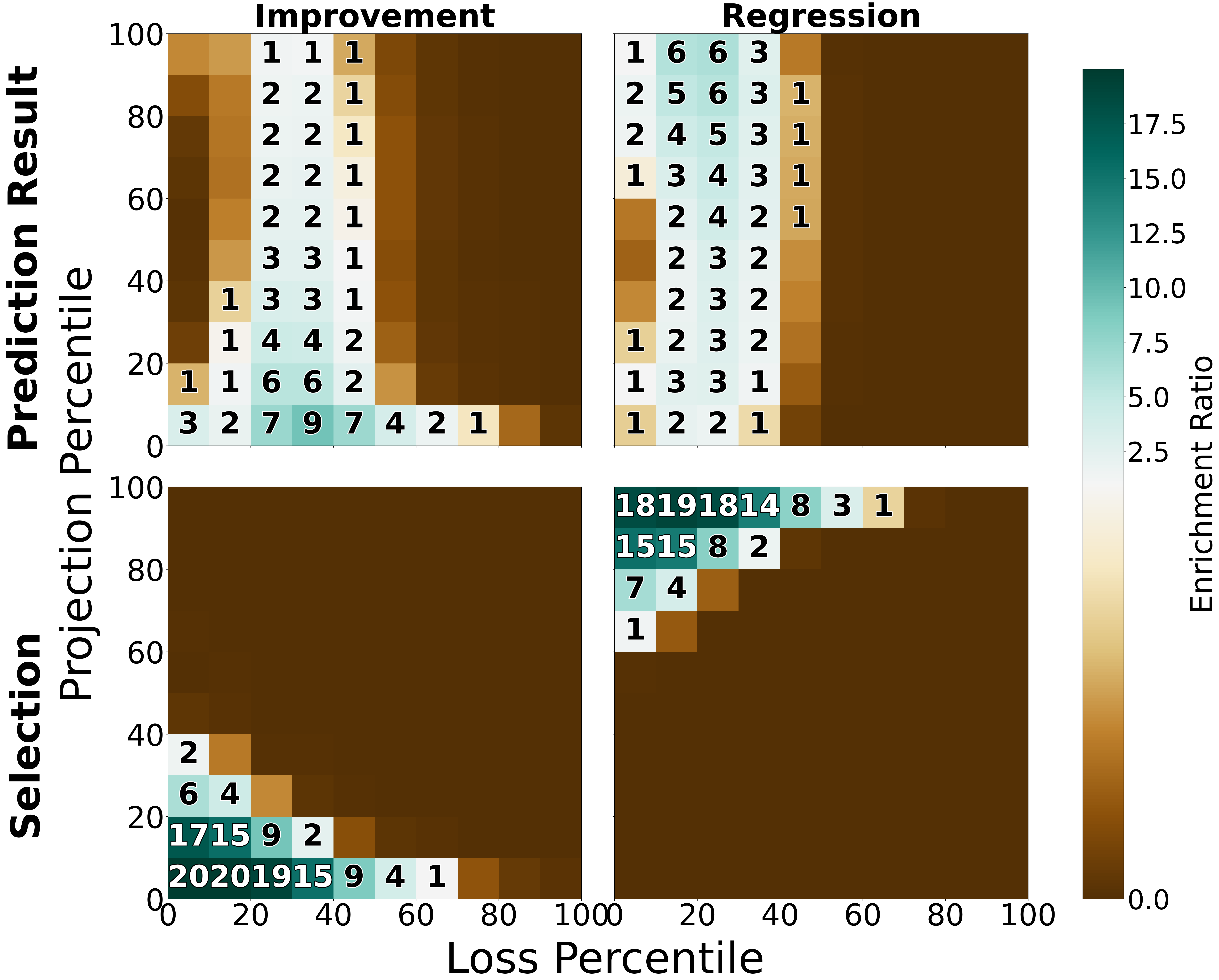}
  \caption{Target-case distributions~(top) and distributions of test cases placed near the head of the GBV-PD ranking~(bottom) for IC~(LeNet5) over original-loss percentile~(x-axis) and projected-delta percentile~(y-axis). Improvement and regression cases occupy distinct hotspots, and the two modes shift the early ranking prefix toward their respective regions. Numbers denote average enrichment ratios per bin.}
  \label{fig:rq2_2}
  \vspace{-0mm}
\end{figure}

Compared with the two evolution-aware references, GBV-PD performs better than MS on most subject-metric combinations. The only cases where MS is significantly better are the two regression-oriented metrics for IC~(ResNet18). Compared with the full-gradient directional reference Grad-PD, GBV-PD is broadly competitive rather than uniformly weaker. It is significantly better in some settings, statistically indistinguishable on BC~(LeNet5), and mixed on IC~(ViT-Tiny) and IC~(LeNet5). This is important because it shows that PCA compression retains most of the useful directional information and, in some cases, filters out high-dimensional gradient components that are not particularly helpful for prioritization.

Figure~\ref{fig:rq2_2} helps explain why the method works. Improvement and regression cases occupy distinct regions in the joint space of the original loss and the projected delta. GBV-PD then shifts the early ranking prefix toward the appropriate hotspot in each mode, rather than simply preferring globally difficult tests. The projected delta provides direction, while the original loss supplies complementary severity information.

\begin{tcolorbox}
    In response to RQ2, GBV-PD is effective as a prioritizer. It consistently places regression and improvement cases earlier than non-directional baselines, usually surpassing MS, and remains broadly competitive with Grad-PD---especially at small budgets, where prioritization matters most.
\end{tcolorbox}

\subsection{RQ3. Efficiency}

\begin{table}[t]
\caption{Operational requirements and measured costs for RQ3 on a representative subject, IC~(ViT-Tiny). Reported times and cache sizes are averages over 10 runs. The 100-update time includes one-time preprocessing and repeated per-update costs.}
\label{table:req}
\centering
\footnotesize
\begin{filecontents*}{table/03_requirement.csv}
Method,loss,neuron,sa,ms,tdpr,gradpd,gbvpd
Original Model Pre-processing,O,X,X,O,O,O,O
Cached Artifact,Loss,None,None,Logits/ ODIN scores,Logits by epochs,Full Gradient,Compact GBV
Cache Size,$\approx$ 50~KB,-,-,$\approx$ 20~MB,$\approx$ 60~MB,$\approx$ 700~MB,$\approx$ 20~MB
Access to Parameter Delta ($\Delta\theta$),X,X,X,X,X,O,O
Evolved Model Train Execution,X,X,X,O,X,X,X
Evolved Model Test Execution,X,O,O,X,X,X,X
Pre-processing Time (s),0.94,0.00,0.00,12.25,0.95,3.82,4.31
Per-update Time (s),0.00,1.06,5.18,20.19,0.00,1.15,0.96
Total Time for 100 Updates (s),0.94,106.50,518.29,2031.09,0.95,118.85,99.82
\end{filecontents*}
\pgfplotstabletypeset[
    col sep=comma,
    every head row/.style={
        before row={\toprule},
        after row=\midrule
    },
    every nth row={1}{before row=\midrule},
    every last row/.style={after row=\bottomrule},
    columns/Method/.style={column type={>{\centering\arraybackslash}m{1.4cm}}, column name=Method, string type},
    columns/loss/.style={column type={>{\centering\arraybackslash}m{0.5cm}}, column name=Loss, string type},
    columns/neuron/.style={column type={>{\centering\arraybackslash}m{0.5cm}}, column name=NC, string type},
    columns/sa/.style={column type={>{\centering\arraybackslash}m{0.5cm}}, column name=SA, string type},
    columns/ms/.style={column type={>{\centering\arraybackslash}m{0.6cm}}, column name=MS, string type},
    columns/tdpr/.style={column type={>{\centering\arraybackslash}m{0.6cm}}, column name=TDPR, string type},
    columns/gradpd/.style={column type={>{\centering\arraybackslash}m{0.95cm}}, column name=Grad-PD, string type},
    columns/gbvpd/.style={column type={>{\centering\arraybackslash}m{0.7cm}}, column name=GBV-PD, string type},
]{table/03_requirement.csv}
\vspace{-0mm}
\end{table}

\begin{figure}[ht]
  \centering
  \vspace{-0mm}
  \includegraphics[width=\linewidth]{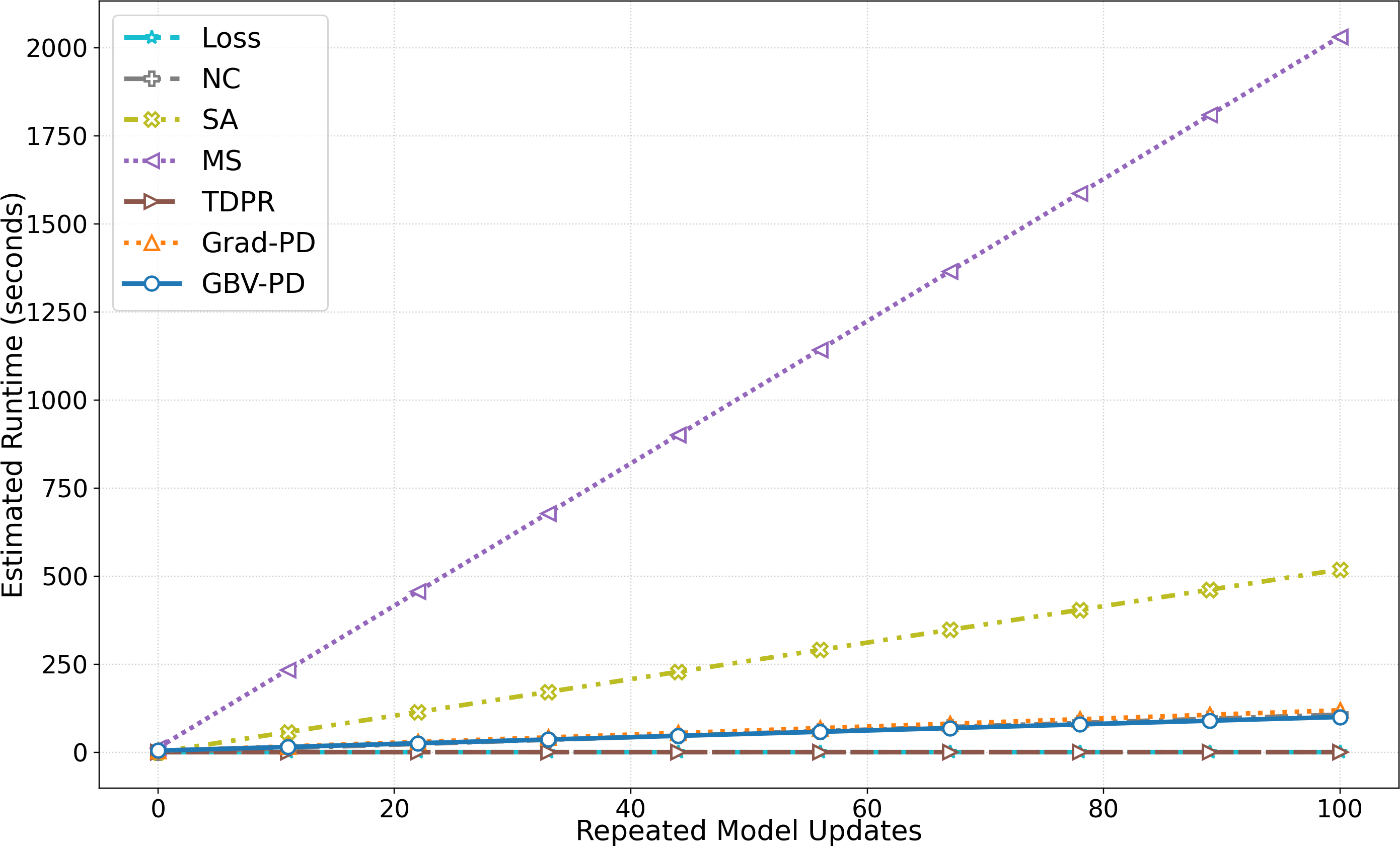}
  \caption{Cumulative estimated runtime over repeated model updates on IC~(ViT-Tiny). SA and NC grow linearly because they rerun the updated model on test cases, whereas MS grows steeply due to its heavy per-update processing. Grad-PD and GBV-PD amortize their preprocessing cost; GBV-PD eventually becomes cheaper than Grad-PD. The estimate assumes reuse of a valid cache built at a fixed reference model.}
  \label{fig:rq3}
  \vspace{-0mm}
\end{figure}

RQ3 asks whether GBV-PD is practical in the repeated-update setting that motivates evolution-aware testing. Even a strong prioritizer is difficult to deploy if it requires large caches or repeatedly runs the updated model over the test set.

We assess operational practicality on a representative subject, IC~(ViT-Tiny), by comparing preprocessing cost, per-update cost, cache size, and whether the updated model must be executed on test cases during prioritization. Reported times and storage are averages over 10 runs.

Table~\ref{table:req} summarizes required artifacts and measured costs, and Figure~\ref{fig:rq3} shows cumulative runtime as updates accumulate. The main comparison is with Grad-PD, which uses the same directional signal without compression. GBV-PD pays slightly more for one-time preprocessing due to PCA, but lowers the per-update cost. After 100 updates, this yields a lower total time. By contrast, NC and SA scale linearly because they rerun the updated model on test cases at each update, whereas MS grows steeply due to its heavy per-update processing.

GBV-PD also reduces storage substantially. As shown in Table~\ref{table:req}, it requires about 20\,MB of cached GBVs, compared with about 700\,MB for full-gradient caching, a reduction of more than 35$\times$. The GBV cache is even smaller than TDPR's logit cache. Crucially, GBV-PD retains update-direction awareness without requiring the test set to execute the updated model during prioritization.

A practical interpretation is that GBV-PD pays once to build a reusable, compact cache, and each subsequent update requires only lightweight projection and ranking. This supports the claim that compression is not only mathematically acceptable but also operationally useful.

\begin{tcolorbox}
    In response to RQ3, GBV-PD achieves operational efficiency. GBV-PD incurs an initial preprocessing cost, but that cost is amortized through reuse of the compact GBV cache. It does not require running the evolved model on the test cases and uses less storage than the Grad-PD. This result supports the claim that GBV-PD meets \textit{Compactness}, \textit{Efficiency}, and  \textit{Automation} requirements.
\end{tcolorbox}

\subsection{RQ4. Robustness}
\label{subsec:rq4}

RQ4 asks whether GBV-PD's gains reflect a stable mechanism or a narrow configuration choice. This matters because an advantage that appears only under one hyperparameter setting would be much less convincing in practice. We vary delta norm $||\Delta\theta||^2_2$, the gradient target layer, the GBV output dimension, the projection ratio--the fraction of $\Delta\theta$ captured by the GBV space--and the parameter $\alpha$ to assess sensitivity to these design choices.

\begin{figure*}[ht]
  \centering
  \vspace{-0mm}
  \includegraphics[width=\linewidth]{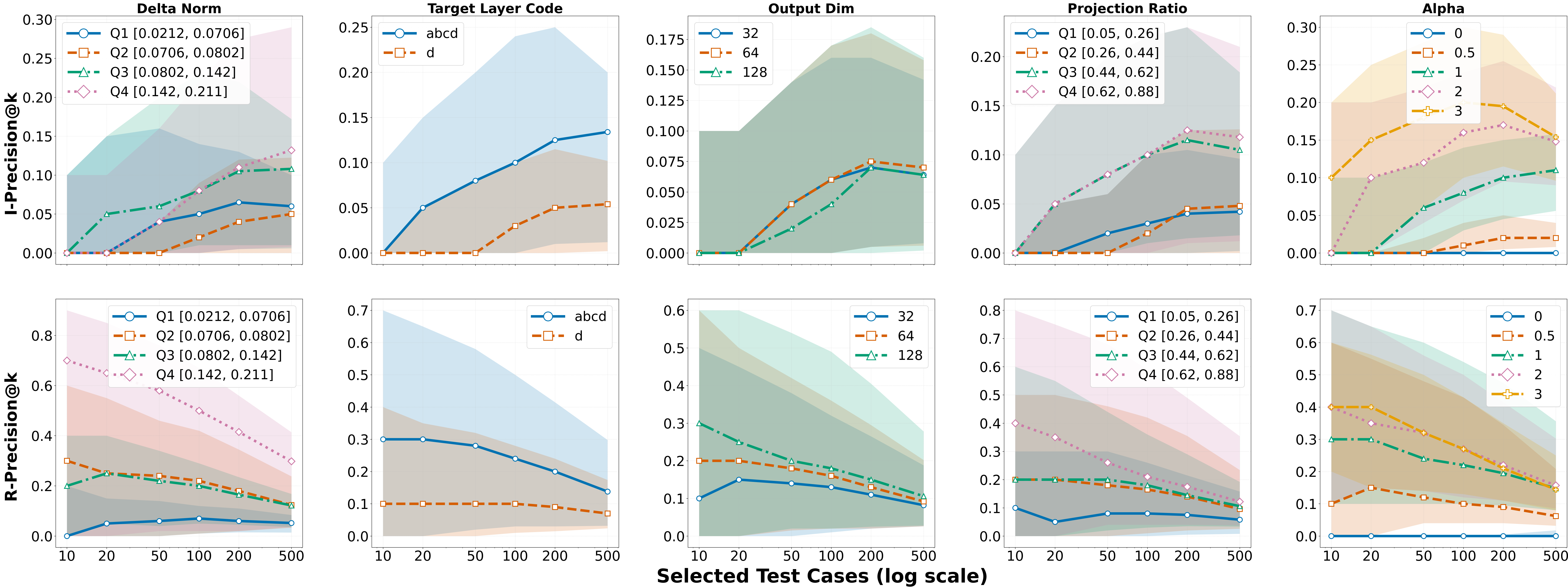}
  \caption{One-factor-at-a-time ablations for GBV-PD on IC~(LeNet5). The strongest effect comes from $\alpha$; multi-layer gradients provide a consistent benefit, and output dimension, projection ratio, and update magnitude mainly have secondary effects. Factors with negligible effects are omitted for readability.}
  \label{fig:rq4}
  \vspace{-0mm}
\end{figure*}

Figure~\ref{fig:rq4} shows the ablation results for IC~(LeNet5); the other tasks exhibit the same qualitative trends. The dominant factor is $\alpha$. When the projected delta loss receives little weight, as with $\alpha=0$ or $0.5$, performance drops markedly, especially at small budgets. Once $\alpha>1$, results improve substantially and are often best at $\alpha=2$ or $3$. This indicates that GBV-PD works because the scoring rule meaningfully emphasizes the update direction, not because it merely reorders tests by their original losses.

This is an important interpretive result. It shows that GBV-PD does not work merely because the original loss already identifies difficult tests. The method improves when the directional term is given real influence, which supports the paper's central claim that update-direction information is the main source of gain.

The next most consistent effect comes from the target layer. Using multiple layers~(\texttt{abcd}) is usually slightly better than using only the final layer~(\texttt{d}), suggesting that update sensitivity is distributed across the network. Other factors, including output dimension, projection ratio, and update magnitude, have visible but secondary effects, while PCA-sampling choices are largely stable.

Overall, GBV-PD is robust across a range of design choices. This is useful from a deployment perspective: the method is tunable, but not fragile. In practice, the most reliable setting is to use multi-layer gradients when feasible and choose $\alpha>1$.

\begin{tcolorbox}
    In response to RQ4, GBV-PD remained broadly robust to changes in behavior vector construction and scoring settings. The strongest performance effects were observed with $\alpha > 1$ and multi-layer gradients.
\end{tcolorbox}
\section{Discussion}
\label{sec:discussion}

\subsection{Operationalization of Behavior Vector Space}

Our work shows that under the same-architecture parameter updates, a gradient-based operationalization of BVS is useful for regression test prioritization. By compressing each test case's gradient into a compact GBV and projecting the update-induced parameter delta onto the same basis, GBV-PD turns post-evolution test case ordering into a directional ranking problem in a compact space. This operationalization shows that a task-specific view of behavior space can be practically useful.

This perspective suggests a pragmatic path for future BVS research. Rather than requiring a full geometric theory from the outset, BVS can be operationalized around the information that matters most for a target testing task. In our case, the key aspect is test-wise sensitivity to parameter updates. The result is a concrete first step toward broader behavior-space-based testing, with future extensions to coverage definition, gap localization, and supplementary test generation.

\subsection{Implications for Software Engineering}

From a software engineering perspective, the immediate contribution of this work is a practical prioritization mechanism for continuously evolving ML components. In MLOps pipelines, models are repeatedly retrained, fine-tuned, or optimized, and rerunning the full test suite after each update is expensive. GBV-PD separates one-time offline computation from lightweight per-update scoring, thereby amortizing the cost of prioritization across updates.

GBV-PD is intended to complement, rather than replace, eventual full-suite execution. In practice, practitioners can execute the top-$k$ prefix of the ranking and inspect the most informative test cases first. Even when a complete rerun is feasible, executing the top-$k$ prefix first can reduce the time to the first actionable regression signal and shorten the feedback loop. Its practical benefit is therefore earlier feedback under a testing budget.

These properties make GBV-PD well aligned with parameter-efficient fine-tuning workflows such as LoRA~\cite{hu2022lora}. In such workflows, models are adapted through repeated, structure-preserving parameter updates rather than structural modifications. As a result, reusable GBV caching and lightweight update-aware scoring can provide a practical mechanism for prioritizing post-update tests without rerunning the full test suite after every adaptation step.

The study also suggests that post-update ML testing benefits from considering regressions and improvements together. A single update can worsen the model's behavior on some test cases while improving it on others, and this asymmetry is easy to miss if one looks only for generally difficult inputs. The directional signal used by GBV-PD helps surface test cases likely to reveal either kind of change. We therefore view improvement-oriented ranking as a complementary diagnostic perspective rather than a replacement for regression-oriented prioritization.

GBV-PD also occupies a useful point in the design space. Black-box or non-directional heuristics do not explicitly encode evolution direction, while full-gradient methods do, but at substantially higher storage and runtime cost across repeated updates. GBV-PD retains update-direction information in a reusable, compact representation, making it attractive for CI/CD-style testing workflows. At the same time, the current operationalization remains at the level of the ML component. Although we motivate the problem through ML-enabled systems, extending the approach to end-to-end system assurance will require integrating non-ML logic, control flow, and, in some domains, closed-loop execution.

\subsection{Limitations}

GBV-PD relies on a local approximation and is therefore best suited to small, repeated updates for which the original and evolved models remain close in parameter space. As updates grow larger, higher-order effects can weaken the relationship between projected delta loss and actual post-update behavior. The method also assumes a fixed architecture with aligned parameters, which makes it natural for retraining, fine-tuning, and other structure-preserving updates, but not for architectural changes such as layer insertion or removal, architecture switching, or expert reconfiguration. Finally, the current formulation assumes supervised settings with labels or task-specific oracles that support loss computation and post hoc regression/improvement labeling. This assumption is less natural in simulation-based evaluation, scenario-based safety testing, or weak-oracle deployment settings.

\section{Threats to Validity}
\label{sec:threats}

\textit{Internal Validity.} The study focuses on same-architecture updates whose magnitudes are small enough for the first-order approximation. Performance may therefore differ for larger or more disruptive updates. Results can also vary with the target gradient layer, PCA sampling strategy, GBV dimensionality, and the weighting parameter $\alpha$. To reduce this risk, we evaluate multiple architectures, update settings, GBV configurations, and $\alpha$ values, and analyze these factors explicitly in RQ4.

Some baselines are sensitive to implementation details and hyperparameter choices. We followed the original designs as closely as possible and kept the evaluation protocol consistent across methods, but alternative implementations or additional tuning could affect absolute results. Moreover, some baselines are defined only for classification settings, so not every comparison is available for every subject. In addition, MetaSel is a test selection method rather than a prioritization method; for comparability under the ranking-based protocol of RQ2, we induced rankings from its per-test scores. This enables a common top-$k$ evaluation setup, but it may not capture every aspect of MetaSel's intended use. Finally, our effectiveness metrics summarize prefix quality across log-scaled budgets, which emphasizes early ranks, and the behavior-cloning labels depend on a predefined threshold to distinguish meaningful changes from negligible numerical differences. Different metric choices or thresholds could shift the quantitative results.

\textit{External Validity.} The evaluation covers image classification and behavior cloning under controlled retraining and fine-tuning settings with aligned parameters. The findings should therefore be interpreted as evidence that GBV-PD is useful in this repeated-update scenario, not as a claim that it applies unchanged to architecture-changing updates, unsupervised settings, or tasks without clear per-test oracles. The current study also focuses on the ML component rather than the full end-to-end system behavior, so additional work is needed to assess how well the approach transfers to system-level testing scenarios involving non-ML logic or closed-loop dynamics. Finally, the time and storage measurements are reported for specific models and hardware; the absolute values may vary across environments, and our cost analysis is intended to highlight relative structural differences among methods rather than universal runtime constants.

\section{Related Work}
\label{sec:related}

Testing of ML-enabled systems has been studied from both model-aware and black-box perspectives. Model-aware approaches such as QuoTe~\cite{chen2023quote}, IDC~\cite{dola2023input}, and NSS~\cite{dola2023input} use internal signals, such as adequacy or neuron sensitivity, to identify informative tests, whereas black-box approaches explore the input or feature space to reveal faults or improve diversity~\cite{zohdinasab2023efficient,aghababaeyan2023black}.

DeepGini~\cite{feng2020deepgini}, FAST~\cite{chen2024fast}, and TDPR~\cite{shen2024prioritizing} rank test inputs using uncertainty, feature selection, or training dynamics, but they assume a static model and therefore do not model how the current update changes individual test cases. Our work differs in evolution-aware design: we prioritize the order of an existing test suite after model evolution, and the interaction between cached per-test behavior and the observed parameter delta drives the ranking.

The closest line of work addresses evolving ML models. Abbasishahkoo et al.~\cite{abbasishahkoo2025metasel} propose MetaSel, a test selection method for fine-tuned DNNs that compares pre-trained and fine-tuned models; You et al.~\cite{you2023regression} use regression fuzzing to generate new inputs that trigger regressions; and Li et al.~\cite{li2025towards} study how to update neural networks while avoiding regressions. These works address selection, generation, or update strategies for evolving models. By contrast, we study prioritization: given an existing test suite and a limited budget, we rank test cases so that the top-$k$ prefix is more likely to reveal post-update regressions or improvements. The repeated-update setting is central to this distinction, because it makes reusable cached representations especially valuable.

Our work is also related to gradient- and representation-based analysis, including TRAK~\cite{park2023trak}, DataInf~\cite{kwon2024datainf}, Dataset Cartography~\cite{swayamdipta2020dataset}, and GRAD-MATCH~\cite{killamsetty2021grad}. These methods use gradients or training dynamics for attribution, data characterization, or subset selection. In contrast, we compress per-test gradients into reusable GBVs and combine them with the current parameter update to support post-update regression test prioritization.

In summary, first-order gradient approximations are also used in influence and attribution methods~\cite{park2023trak,kwon2024datainf}, but their objective is typically to estimate data influence or attribution rather than regression-test prioritization. GBV-PD instead treats the observed parameter delta as an explicit evolution direction and converts signed, test-wise update effects into a budget-aware ranking of an existing test suite. It also differs from selection, generation, and update-repair techniques for evolving models~\cite{abbasishahkoo2025metasel,li2025towards,you2023regression} in both its output—a full prioritization—and its online requirement, which does not involve executing the evolved model on the test suite.
\section{Conclusion}
\label{sec:conclusion}

This paper studied evolution-aware regression test prioritization for ML components under parameter updates within the same architecture. We introduced GBV-PD, which builds a compact behavior vector from per-test gradients and combines it with the parameter delta to estimate which tests are most likely to regress or improve, without executing the evolved model on the full test set. Across image classification and behavior cloning, GBV-PD preserved the ranking of post-update loss changes, consistently outperformed non-directional baselines, and remained competitive with the full-gradient reference Grad-PD. It also offered a more practical cost profile for repeated updates via reusable GBV caching. These results show that behavior-space ideas can be operationalized for real regression testing problems.

Future work includes extending GBV-PD to broader architecture-changing updates and to weak-or-no-oracle settings such as reinforcement learning and simulation-based testing, as well as developing BVS further through analyses of how prioritization outcomes reflect changes in behavioral regions and through prospective BVS-based coverage criteria. We hope that this work lays the groundwork for practical support for continuous verification and post-evolution analysis, with the potential to improve the reliability and maintainability of continuously evolving ML-enabled systems.

\section*{Data Availability}
Our replication package~\cite{replication} is publicly available and additionally contains full experiment results on Figshare.

\begin{acks}
This work was partly supported by the Institute of Information \& Communications Technology Planning \& Evaluation(IITP)-ITRC~(Information Technology Research Center) grant funded by the Korea government(MSIT) (2026-RS-2020-II201795, 10\%) and Institute of Information \& communications Technology Planning \& Evaluation(IITP) grant funded by the Korea government~(MSIT)((RS-2025-02218761, 50\%), (RS-2024-00406245, 15\%)), and Samsung Electronics(25\%).
\end{acks}

\bibliographystyle{ACM-Reference-Format}
\bibliography{reference}

\end{document}